\documentclass[usenatbib,letterpaper]{mn2e}
\usepackage{amsmath}
\usepackage{amssymb}
\usepackage{amsbsy}
\usepackage{graphicx}
\usepackage{ifthen}
\usepackage{rotating}
\usepackage{calc}
\usepackage[colorlinks=true]{hyperref}

\newboolean{lores}
\setboolean{lores}{true}

\newcounter{B07Done}
\def\Bett{\ifthenelse{\equal{\arabic{B07Done}}{0}}{\citeauthor{bett_spin_2007}~(\citeyear{bett_spin_2007}; hereafter B07)\setcounter{B07Done}{1}}{B07}}
\def\Bettp{\ifthenelse{\equal{\arabic{B07Done}}{0}}{(\citeauthor{bett_spin_2007}~\citeyear{bett_spin_2007}; hereafter B07)\setcounter{B07Done}{1}}{(B07)}}
\def\Betta{\ifthenelse{\equal{\arabic{B07Done}}{0}}{\citeauthor{bett_spin_2007}~\citeyear{bett_spin_2007}; hereafter B07\setcounter{B07Done}{1}}{B07}}

\newcounter{MCMCDone}
\def\MCMC{\ifthenelse{\equal{\arabic{MCMCDone}}{0}}{Markov Chain Monte Carlo (MCMC)\setcounter{MCMCDone}{1}}{MCMC}}

\newcounter{SAMDone}
\def\SAM{\ifthenelse{\equal{\arabic{SAMDone}}{0}}{semi-analytic model (SAM)\setcounter{SAMDone}{1}}{SAM}}
\def\SAMs{\ifthenelse{\equal{\arabic{SAMDone}}{0}}{semi-analytic models (SAMs)\setcounter{SAMDone}{1}}{SAMs}}

\title[The Noise-Free Distribution of Spin Parameters]{Constraining the Noise-Free Distribution of Halo Spin Parameters}
\author[Andrew J. Benson]{Andrew J. Benson\\
Carnegie Observatories, 813 Santa Barbara Street, Pasadena, CA 91101, USA;\\Kavli Institute for Theoretical Physics, University of California, Santa Barbara.}

\begin{document}

\maketitle

\begin{abstract}
Any measurement made using an N-body simulation is subject to noise due to the finite number of particles used to sample the dark matter distribution function, and the lack of structure below the simulation resolution. This noise can be particularly significant when attempting to measure intrinsically small quantities, such as halo spin. In this work we develop a model to describe the effects of particle noise on halo spin parameters. This model is calibrated using N-body simulations in which the particle noise can be treated as a Poisson process on the underlying dark matter distribution function, and we demonstrate that this calibrated model reproduces measurements of halo spin parameter error distributions previously measured in N-body convergence studies. Utilizing this model, along with previous measurements of the distribution of halo spin parameters in N-body simulations, we place constraints on the noise-free distribution of halo spins. We find that the noise-free median spin is 3\% lower than that measured directly from the N-body simulation, corresponding to a shift of approximately 40 times the statistical uncertainty in this measurement arising purely from halo counting statistics. We also show that measurement of the spin of an individual halo to 10\% precision requires at least $4\times 10^4$ particles in the halo---for halos containing 200 particles the fractional error on spins measured for individual halos is of order unity.
\end{abstract}


\section{Introduction}\label{sec:introduction}

The angular momentum content of dark matter halos, which arises through tidal torques acting during halo collapse \citep{hoyle_origin_1949,peebles_origin_1969,doroshkevich_spatial_1970,white_angular_1984,barnes_angular_1987,porciani_testing_2002}, and which is typically characterized by the dimensionless spin parameter, $\lambda$ \citep{peebles_origin_1969}, is of key importance for the study of galaxy formation \citep{benson_galaxy_2010}. As demonstrated by \citeauthor{fall_formation_1980}~(\citeyear{fall_formation_1980}; see also \citealt{mo_formation_1998}), the spin parameter (to the extent that baryonic material angular momentum traces that of the dark matter) directly determines the angular momentum content of galaxies and, therefore, their sizes. The distribution of spin parameter values is therefore of interest as it directly determines the spread in galaxy sizes at fixed mass. To obtain an accurate measure of this distribution the usual approach is to measure it directly from N-body simulations, and this has been done by several authors \citep{cole_structure_1996,bett_spin_2007,gottlober_shape_2007,maccio_concentration_2007,zhang_spin_2009,lee_properties_2016,rodriguez-puebla_halo_2016,zjupa_angular_2017}, finding median spins of $\lambda\approx 0.038$ \Bettp.

In an N-body simulation, a set of particles are used as both tracers of the dark matter distribution function, and as carriers of the source of gravitational potential (i.e. mass). In their role as tracers of the dark matter density field a simple assumption is that the particles represent a Poisson process, that is, they are independently drawn random locations within the simulation volume with a distribution proportional to the dark matter density field at each position. This assumption may be challenged on the basis that the dark matter particles do not start out as a Poisson process (typically the initial distribution of particle positions is derived from a glass \citep{bagla_cosmological_1997}, which is not a Poisson process), and because the particles are also the source of gravitational potential, which could lead a break down of the assumption that particle positions are independent. However, in Appendix~\ref{sec:nbodyStats} we demonstrate the the actual particle distribution in cosmological N-body simulations is consistent with a Poisson process.

Given this understanding of what the particles represent, it is clear that a measurement of \emph{any} property in a simulation is subject to noise, arising from the discrete sampling of the true dark matter distribution function. Furthermore, as shown by \cite{trenti_how_2010}, additional noise arises due to the lack of structure below the resolution limit of the simulation. In many cases, such as a measurement of mass \citep{benson_mass_2016}, this noise will lead to variance in the measured quantity. In some cases, such as the one to be discussed in this work, it can also lead to a bias. In either case, it is important to understand that any measured distribution of properties of a class of objects (e.g. halos) in an N-body simulation, is a measurement of the true distribution convolved with some distribution of noise arising from the finite number of particles.

While this was certainly not a significant issue for early N-body simulations---in which, for example, the small number of halos lead to statistical fluctuations in measured distributions which far exceeded the effects of particle noise---we will show in this work that even the large volumes and high resolution of simulations from a decade ago result in finite particle number effects which are very significant, and lead to biases in measured parameters which far exceed the uncertainties arising from halo number statistics if ignored.

In \cite{benson_mass_2016} we showed how these finite particle number effects affect measurements of the dark matter halo mass function. In that case the effects, while significant compared to the measurement uncertainties, are small in an absolute sense. In the current work, we identify a case where finite particle number effects are substantial both compared to the statistical uncertainties, but also in an absolute sense.

Specifically, we consider halo spin parameters, defined as \citep{peebles_origin_1969}:
\begin{equation}
 \lambda = \frac{J |E|^{1/2}}{{\rm G} M^{5/2}},
 \label{eq:spin}
\end{equation}
where $J$ is the magnitude of the halo's angular momentum, $E$ is the energy of the halo (consisting of both gravitational potential and kinetic energy), and $M$ is the halo mass. The spin parameter measures the net angular momentum of a halo and is defined such that a spin parameter of $\lambda\sim 1$ would be obtained for a halo supported by rotation. In fact measured median spin parameters are $\tilde{\lambda}\approx 0.04$ \Bettp, indicating that halos are primarily pressure supported, with individual particles distributed on roughly isotropic orbits. As such, any measurement of spin is attempting to measure a small rotation component against a large background of ``noise'' (i.e. the randomly oriented pressure supported orbits). This is a case where finite particle number effects may therefore be expected to significantly affect the measurement. As discussed by \Bett\ (and as we will describe in more detail below) this noise also leads to a biased measurement.

To make an order of magnitude estimate of the effects of particle noise we can consider a simple model. We consider the distribution of the components of individual particle angular momenta along the axis of the true angular momentum of the halo\footnote{As we are interested here in the fractional error in the spin parameter, we can equally well work directly with the angular momentum vector since $\lambda\propto J$.} to be described by a normal distribution with mean $\tilde{\lambda}$, and a dispersion, $\sigma_\lambda$, of order unity (corresponding to the velocity dispersion of the halo). If we sample $N$ values from this distribution, the mean can be estimated with an uncertainty of approximately $\sigma_\lambda/\sqrt{N}$. In order to have a reliable measurement we therefore require, $\sigma_\lambda/\sqrt{N} \ll \tilde{\lambda}$. This implies a required $N\gg (\sigma_\lambda/\lambda)^2\sim 10^3$. Similarly, a measurement of the spin in an individual halo with a precision of 10\% would require at least $10^5$ particles. As we will show later, a detailed calculation (using the correct statistics for the noise) validates this order of magnitude estimate---a 10\% measurement typically requires around $4\times 10^4$ particles---it is clear that spins of individual halos can be trusted only for very well resolved halos. Of course, measurement of properties of populations of halos (such as the mean spin) can be determined more accurately, but the spin distribution will be unavoidably affected by finite particle number noise.

We suggest that the correct approach to measuring \emph{any} property from an N-body simulation is to forward model the effects of particle noise on a model for the  true (noise-free) distribution, and then constrain the parameters of that true distribution from the measured distribution. In this way\footnote{Assuming that the forward model of the effects of finite particle number is sufficiently accurate.}, the constrained parameters should be unbiased, stable against changes in resolution and indicative of the true properties of dark matter halos, rather than those measured with a finite number of particles.

In this work, we develop a model for the effects of finite particle number noise on the measured spins of dark matter halos, and use that model, along with the data from \Bett\ to constrain parameters of the intrinsic, noise-free distribution of dark matter halo spins. The remainder of this work is organized as follows. In \S\ref{sec:methods} we describe our model for the effects of particle noise on halo spins, and describe how we apply this to the intrinsic distribution of spin parameters to derive the distribution actually measured in an N-body simulation. In \S\ref{sec:results} we present results for the intrinsic distribution of spin parameters, and finally in \S\ref{sec:discussion} we discuss the implications of these results. Additionally, in Appendix~\ref{sec:nbodyStats} we perform N-body experiments to validate the assumption that N-body simulation particles are consistent with being a Poisson process.

\section{Methods}\label{sec:methods}

In this section we develop a simple model to describe the effects of finite particle number noise on the measurement of spin parameters in halos extracted from N-body simulations. \cite{trenti_how_2010} have provided a simple model for the distribution of halo masses arising from noisy N-body measurements, and we will make use of their model in our calculations. However, before doing so we first conduct a more limited experiment in N-body particle noise---neglecting the effects of changing resolution which \cite{trenti_how_2010} examined---to both test and provide a calibration sample for our model of the spin parameter noise distribution, and to justify the assumption that N-body particles represent a Poisson process applied to the underlying dark matter distribution function.

\subsection{Monte Carlo Estimates of Errors on Halo Spins}\label{sec:monteCarlo}

To make a direct estimate of the errors on the measured spin parameters of N-body halos we use a Monte Carlo bootstrapping procedure on a sample of halos extracted from the Millennium Simulation \citep{springel_simulations_2005} spanning a range of around $10^2$ to $10^5$ in the number of particles contained within the halo. While this approach will miss a contribution to the errors from the lack of structure below the resolution limit (see \citealt{trenti_how_2010}) it is sufficient for the purpose of calibrating our model for halo spin parameter errors providing we correctly account for this missing component of the error budget. We follow the procedure of \cite{benson_mass_2016} to bootstrap resample the particles from each halo. Briefly, we extract all particles within a cubic region of length 6~Mpc/$h$ (sufficient to more than capture the entirety of the halo plus a significant region around it) centred at the halo centre of mass. Then, for each particle $i$, we draw a weight $w_i$ from a Poisson distribution with mean $\mu$, to produce a new sample of particles (containing, on average, $\mu N$ particles). We then apply the friends-of-friends algorithm \citep{davis_evolution_1985} to identify which particles in this new sample are judged to belong to the halo. Finally, we measure the mass, angular momentum, and energy of these halo particles and compute the corresponding spin parameter using equation~(\ref{eq:spin}). This resampling is repeated a large number of times for each halo. The mean and variance of the set of measured spin parameters is then recorded for each halo. Furthermore, we repeat this process for several values of $\mu$, starting from $\mu=1$ and decreasing by factors of 2 until $\mu N < 100$.

Since we sample particles with replacement, some particles may be sampled more than once (i.e. the weights $w_i$ can be greater than 1). When applying the friends-of-friends algorithm to this set of sampled particles it is necessary to account for this feature---if we count such over-sampled particles only once the mean inter-particle spacing would be increased, thereby changing the isodensity threshold selected by the friends-of-friends linking length. Therefore, we classify friends of particle $i$ as all particles $j$ within a distance $l(w_i^{1/3}+w_j^{1/3})/2$ of particle $i$, where $l$ is the original linking length. This ensures that the correct isodensity contour is recovered. Without this correction halo masses would be biased low. For example, consider the largest halo in our sample, which consists of 56,682 particles in the original simulation. If we do not account for multiple samplings of particles in this way the mean number of particles recovered is approximately 51,000 (around 10\% low) when $\mu=1$, while including the above correction results in a mean number of particles consistent with that in the original simulation.

Multiple samplings also affect measurements of the potential energy of each halo. Normally, the potential energy, $W$, is computed from the particles using
\begin{equation}
 W = - {N \over  N - 1} {\rm G} m^2 \sum_{i=1}^N \sum_{j=1, j\ne i}^N {s(r_{ij}) \over r_{ij}},
\end{equation}
where $m$ is the mass per particle, $r_{ij}$ is the separation of particles $i$ and $j$, $s(r)$ accounts for the deviation from a Newtonian potential due to the inclusion of softening in the N-body simulation, and the $N/(N-1)$ corrects for the lack of self-interaction energy. In our case we write
\begin{equation}
 W = - {N \over  N - 1 - \mu} {\rm G} m^2 \sum_{i=1}^N \sum_{j=1, j\ne i}^N {w_i w_j s(r_{ij}) \over r_{ij}},
\end{equation}
where the change in the first term reflects the fact that some particles are sampled more than once, with $\mu$ being the mean sampling rate, and the particle mass $m$ is increased appropriately (i.e. by $1/\mu$).
% If there are $N^\prime$ particles in our sample, each with a weight $w_i$, then the sum over these weights is $N^{\prime 2} \langle w\rangle^2 - N^\prime \langle w^2\rangle$. By definition, $\langle w \rangle = \mu$, and for a Poisson distribution $\langle w^2\rangle = \sum_{j=1}^\infty j^2 P(j) = \mu + \mu^2$, such that the sum over the weights can be written $N^{\prime 2} \mu^2 - ^\prime ( \mu + \mu^2 )$. Defining $N = \mu N^\prime$ the summation can be written $N (N - 1 - \mu)$. The correction term is therefore $N^2 / N (N-1-\mu) = N/(N-1-\mu)$.

Since we wish to compare our model for the error distribution of N-body halo spin measurements with the results of \Bett\ we follow their approach and exclude halos with $Q>0.5$ where $Q = |2T/U+1|$, $T$ is the kinetic energy of the halo and $U$ is its gravitational potential energy. This excludes halos which are far from virial equilibrium (such as recently merged halos for example). Furthermore, when computing the potential energy of halos \Bett\ used a random sample of 1000 particles to estimate the potential energy in halos containing more than 1000 particles. To compare to their results we compute a potential energy estimate in this way, but also compute potential energies using the full complement of particles in all halos for comparison.

This Monte Carlo approach assumes that repeated Poisson sampling of particles from an individual N-body simulation is equivalent to repeating an N-body simulation many times, each with identical initial density field, but with those initial conditions sampled by different random distributions of particles. This assumption is experimentally verified in Appendix~\ref{sec:nbodyStats}.

\subsection{Model for Spin Errors}

\subsubsection{Model}

To construct a model for the error on halo spin due to particle noise, we assume that halos are spherically symmetric, with a velocity distribution function that is isotropic plus a small degree of rotation (corresponding to the spin). We consider the measured spin vector, $\boldsymbol{\lambda}_{\rm meas}$, to be the sum of true and noise spin vectors, that is
\begin{equation}
\boldsymbol{\lambda}_{\rm meas} = \boldsymbol{\lambda}_{\rm true} + \boldsymbol{\lambda}_{\rm noise}.
\end{equation}
The angular momentum vector of the halo can also be written as
\begin{equation}
\mathbf{J}_{\rm meas} = \mathbf{J}_{\rm true} + \mathbf{J}_{\rm noise}.
\end{equation}
We model the effects of particle noise on the angular momentum vector by considering a 3D random walk. We consider the specific angular momentum of each particle to be drawn from a distribution with mean value $\mathbf{J}_{\rm true}/M$ (i.e. on average each particle has a specific angular momentum equal to the angular momentum of the entire halo by definition) with dispersion of $\sigma_j$. After $N$ steps in a 3D random walk we therefore expect that the mean squared displacement from the starting position in angular momentum space is $\sigma^2_{\rm J} = N m_{\rm p}^2 \sigma_j^2$. We write this as $\sigma^2_{\rm J} = \gamma^2 J^2_{\rm s} / N$, where $J_{\rm s}={\rm G} M^{5/2}/|E|^{1/2}$, and $\gamma = N m_{\rm p} \sigma_j / J_{\rm S} \equiv M \sigma_j / J_{\rm s}$. Equivalently, in terms of the spin parameter, $\sigma^2_\lambda = \gamma^2 / N$. As such, the contribution to the error from the random walk of the angular momentum vector is independent of spin, and depends only on particle number. We will treat $\gamma$ (which we expect to be of order unity) as a free parameter to be adjusted to best match the results of our Monte Carlo experiment.

Furthermore, if we model the distribution of offsets around the true angular momentum vector in each dimension as a normal distribution (which is reasonable by the central limit theorem as it is the sum of a large number of random variables), then the scaled, squared magnitude of the vector, $|J_{\rm meas}/\sigma_{\rm J}|^2$, will follow a non-central $\chi^2$ distribution with 3 degrees of freedom. Specifically, the non-centrality parameter is $W = J_{\rm true}^2/\sigma_{\rm J}^2$. This distribution describes both the bias introduced into the measured spin by the noise vector (as discussed by \Betta, this bias arises because for a isotropically distributed noise vector added to a signal vector, the vector sum is more likely to lie further from the origin than the true vector), and the dispersion around the intrinsic spin.

Due to the definition of the spin parameter (see equation~\ref{eq:spin}) the error in the spin receives contributions from not only the error in angular momentum, but also the errors in the mass, and energy of the halo. We assume that the energy can be written as
\begin{equation}
 E = \alpha \frac{{\rm G} M^2}{r},
\end{equation}
where $r$ is the radius of the halo, and $\alpha$ is some form factor which will depend on the structure (density profile) of the halo. If we further assume that fluctuations in the radius of the halo, $\Delta r$, are related to fluctuations in the mass, $\Delta M$, by $\Delta r / r = \beta (\rho_{\rm i}/3\rho_{\rm s}) \Delta M/M$ where $\rho_{\rm i}$ is the mean interior density of the halo, $\rho_{\rm s}$ is the density at the surface of the halo, and $\beta$ is some factor of order unity which will depend on the nature of group finding, then this contribution to the error on spin will be given by
\begin{equation}
 \sigma_\lambda = \left[ \frac{5}{2}+\left(1+\frac{\beta}{2}\frac{\rho_{\rm i}}{3\rho_{\rm s}}\right) \right] \lambda  \frac{\sigma_N}{N},
\end{equation}
where we have assumed $\alpha$ to be constant (i.e. to not change as the number of particles in the halo fluctuates), and we have assumed that the various error terms add directly (rather than in quadrature) as they all arise from the same underlying fluctuation in particle number and so are correlated. We assume that these errors can be modeled as a log-normal distribution, and will adjust the value of $\beta$ to match errors measured using our Monte Carlo approach. Note that the error contributions arising from uncertainties in the mass and energy of the halo are proportional to the spin parameter of the halo.

The final distribution of $\lambda_{\rm meas}$ for a halo of given mass and true spin is modeled as the product distribution of the non-central $\chi^2$ and log-normal distributions used to describe the spin-independent and spin-dependent contributions to the error respectively.

\subsubsection{Comparison with Monte Carlo Estimates}

Our model for the error distribution of halo spins measured from N-body simulations has two parameters, $\beta$ and $\gamma$, which we expect to be of order unity, but for which we do not know precise values. To constrain these parameters we make use of the Monte Carlo estimates of the variance of measured halo spins as described in \S\ref{sec:monteCarlo}. Specifically, we seek the values of $\beta$ and $\gamma$ which optimize (in a $\chi^2$ sense) the match between our model and trends of variance in halo spin with both particle number and with measured spin (this second trend being important to ensure that our model captures the bias introduced into measured spins by particle noise). We find that values of $\beta=0.4025$ and $\gamma = 0.4175$ give an adequate description of the Monte Carlo experiment results, and, as expected, are of order unity.

\begin{figure}
  \includegraphics[width=85mm]{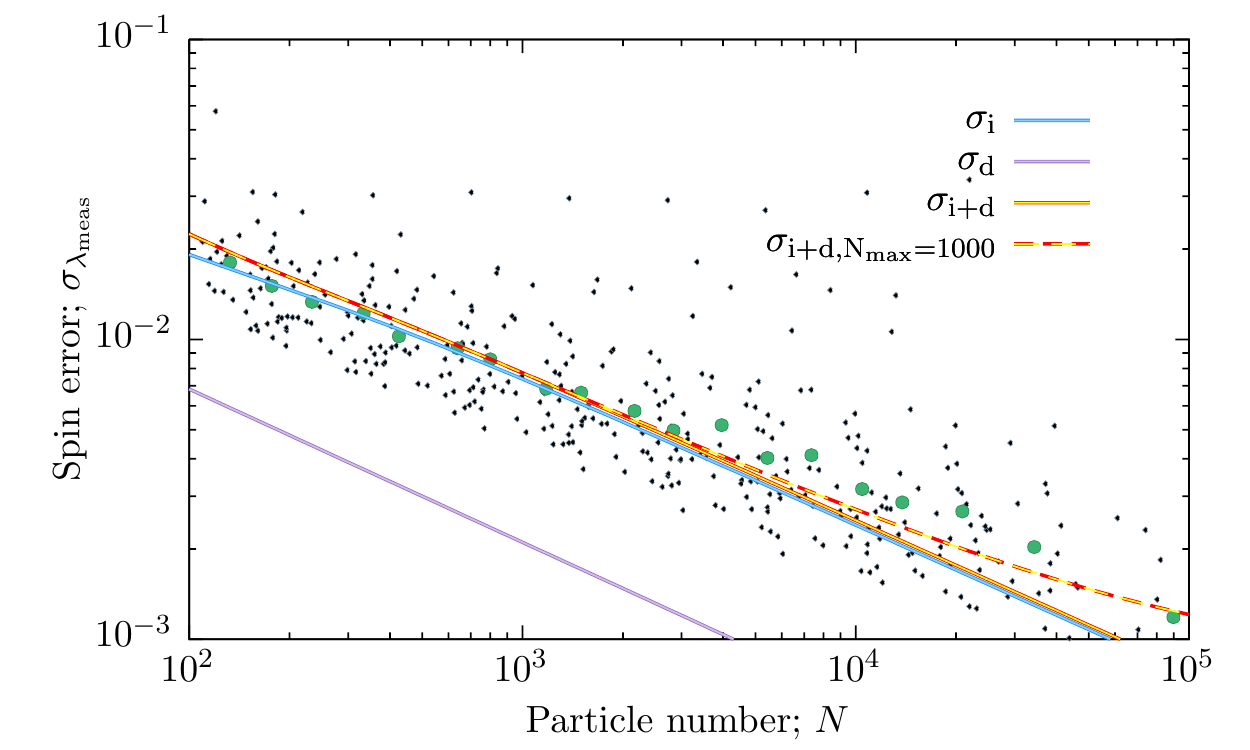}
  \caption{The measured root-variance in halo spin as a function of halo particle count for the Millennium simulation determined from Monte Carlo experiments. Small black points show results obtained by our bootstrap procedure using the entire particle complement to compute the halo potential energy, while the green circles show binned means of the individual halo results. The violet and blue lines show the spin-dependent ($\sigma_{\rm d}$) and spin-independent ($\sigma_{\rm i}$) contributions to the error respectively (in the spin-dependent case we assume $\lambda=0.03$) as predicted by our model, while the yellow line shows the total error ($\sigma_{\rm i+d}$; resulting from the product distribution of the spin-dependent and spin-independent contributions). The dashed yellow-red line shows the total error when we boost the error in energy to account for the 1000-particle limit used by \protect\Bett\ in computing the potential energy of halos.}
  \label{fig:spinErrorVsMass}
\end{figure}

Figure~\ref{fig:spinErrorVsMass} shows the measured root-variance in halo spin as a function of halo particle count for the Millennium simulation determined from Monte Carlo experiments. Small black points show results obtained by our bootstrap procedure using the entire particle complement to compute the halo potential energy, while the green circles show binned means of the individual halo results. The violet and blue lines show the spin-dependent ($\sigma_{\rm d}$) and spin-independent ($\sigma_{\rm i}$) contributions to the error respectively (in the spin-dependent case we assume $\lambda=0.03$) as predicted by our model, while the yellow line shows the total error ($\sigma_{\rm i+d}$; resulting from the product distribution of the spin-dependent and spin-independent contributions). The dashed yellow-red line shows the total error when we boost the error in energy to account for the 1000-particle limit used by \Bett\ in computing the potential energy of halos.

\begin{figure}
  \includegraphics[width=85mm]{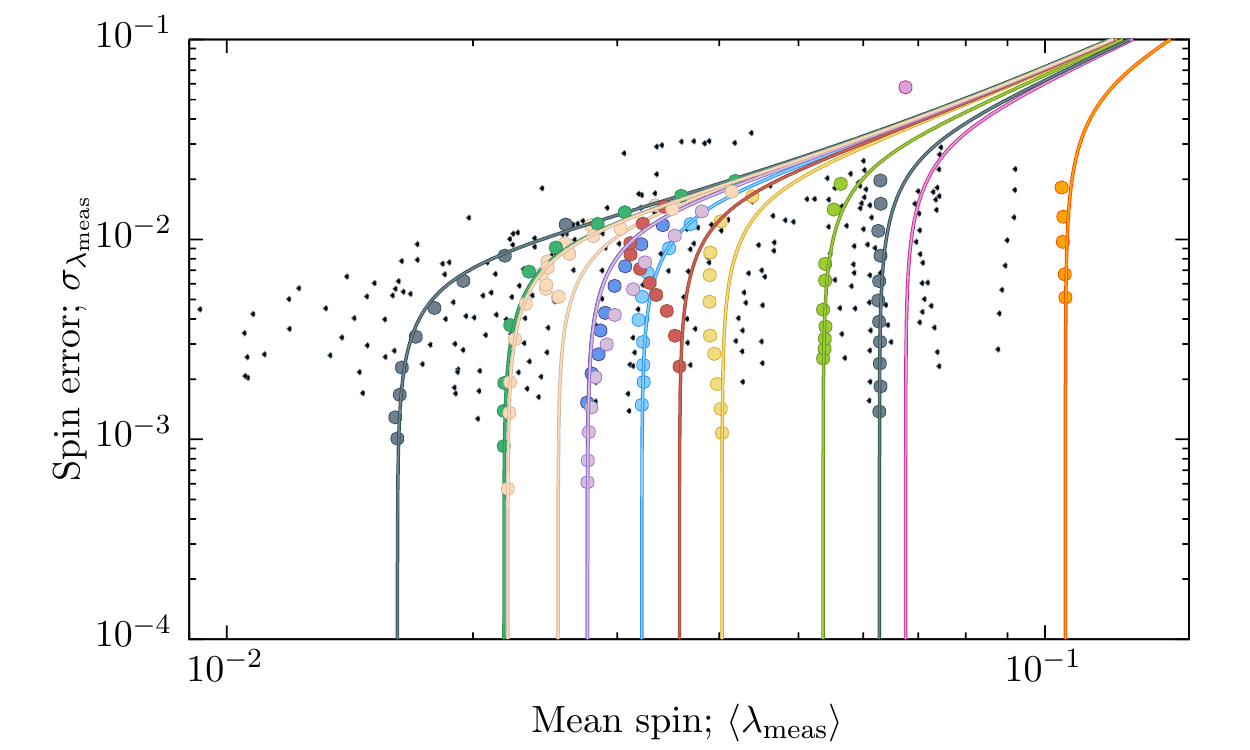}
  \caption{The measured root-variance of halo spins, $\sigma_{\lambda_{\rm meas}}$, as a function of the mean measured spin, $\langle \lambda_{\rm meas} \rangle$. Points show results for individual halos for which particles were sampled at various rates, $\mu$. For several massive halos we plot coloured points to show how the mean measured spin changes with sampling rate (and, therefore, with the total number of particles in the halo). For these massive halos we plot (using lines of the corresponding colour) the trajectories predicted by our simple model, using the mean measured spin at $\mu=1$ as our estimate of the true spin.}
  \label{fig:spinErrorVsSpin}
\end{figure}

Figure~\ref{fig:spinErrorVsSpin} shows the root-variance of halo spins from our Monte Carlo experiments as a function of the mean spin measured in those same experiments. Points show results for individual halos from the Millennium Simulation. For halos containing a sufficiently large number of particles in the original simulation ($N>40,000$) we plot the results as larger coloured points, such that points of a given colour correspond to the same halo, measured with different sampling rates $\mu$. For these large halos we plot the expected trajectory according to our model for the error distribution on N-body halo spins. For the spin-dependent term in these trajectories we set the true spin equal to the measured spin of each halo for $\mu=1$---the fact that the measured trajectories are close to vertical at $\mu=1$ (i.e. where $\sigma_{\lambda_{\rm meas}}$ is smallest) indicates that this is in fact a good estimate of the true spin.

Overall, our simple, analytic model for the distribution of spins measured from N-body halos performs well. In Figure~\ref{fig:spinErrorVsMass} we see that the spin-dependent contribution to the error is always sub-dominant\footnote{Of course, since it is spin-dependent, for very high spin halos it will become comparable to the spin-independent term.}. The model reproduces the trend in mass measured in our Monte Carlo experiment. There is a noticeable under-prediction of the measured root-variance at $N\gtrsim 10^4$ by around 40\%, indicating that some of our model assumptions may be failing in this regime (for example, the small black points in Figure~\ref{fig:spinErrorVsMass} show many large-error outliers which typically correspond to merging halos that are distinctly non-spherical and bias the mean spin error high). Nevertheless, since this is the regime where the error is small we do not expect this small discrepancy to significantly affect the results that we obtain below.

Figure~\ref{fig:spinErrorVsMass} also clearly shows that achieving a 10\% precision measurement of halo spin in an individual halo requires at least $N=10^4$ particles in that halo for typical spins of around $0.03$, while the fractional error in halo spin reaches order unity for $N=100$. This is better than expected from the order-of-magnitude argument made in \S\ref{sec:introduction}, (which predicted $N=10^5$ would be required for a 10\% precision measurement---the difference is that we have now correctly included the various order unity factors and calibrated to real, cosmological halos)---but this is still a large number and the error budget in these simulations lacks any contribution arising from missing structure below the resolution limit \citep{trenti_how_2010}.

In Figure~\ref{fig:spinErrorVsSpin} we see that our simple model accurately reproduces the trend of measured spin root-variance vs. mean measured spin. The trajectories can be understood as follows. When a halo is sampled by a large number of particles, the error on the measured spin is small, and mean measured spin therefore lies close to the true spin. As the number of particles is reduced (as happens when we decrease $\mu$ in our Monte Carlo experiment) the error on the measured spin increases. As this error becomes a significant fraction of the true spin, the bias effect\footnote{Briefly, when adding an isotropically distributed noise vector to the true spin vector, more than half of the resulting summed vectors will have magnitude exceeding that of the true spin vector, leading to a bias in the measured spin. As the magnitude of the noise vector grows the measured spin will approach the magnitude of that noise vector.} discussed by \Bett\ becomes important, and the mean measured spin begins to increase. Our model captures the form of these trajectories well in most cases. There are some instances (notably the trajectory shown by dark red points in Figure~\ref{fig:spinErrorVsSpin}) for which the trajectory found from Monte Carlo experiments shows a more complicated behaviour, with the mean spin initially decreasing as the sampling rate $\mu$ is decreased. These instances appear to be associated with halos that violate some of the assumptions of our model (notably spherical symmetry), in some instances due to the way in which the friends-of-friends algorithm links together nearby overdense regions.

\subsection{Comparison with Trenti et al. (2010)}

\cite{trenti_how_2010} demonstrate that the error in measurements of halo properties from N-body simulations is dominated by effects other than the Poisson process sampling of the dark matter distribution function---specifically differences in the gravitational forces and integration arising from the lack of smaller scale structure which should be present if infinite resolution were available. We will therefore use their model for errors in halo masses when constraining the distribution of halo spin parameters. \cite{trenti_how_2010} demonstrate that the fractional error in the number of particles in a halo can be described by a simple relation
\begin{equation}
 \epsilon(N) \approx 0.15 \left({N \over 1000}\right)^{-1/3}.
 \label{eq:errorTrenti}
\end{equation}
In the majority of the cases to which this relation was fit, \cite{trenti_how_2010} compared simulations which differed in the total number of particles used by a factor 8. The errors they measure in the ratio of halo masses are therefore actually the combined errors from those two simulations. Assuming the simulations to be independent, this means that the normalization of eqn.~(\ref{eq:errorTrenti}) is overestimated by a factor of $\sqrt{5/4}$. We therefore adopt the same relation but with a normalization of $0.135$ instead of $0.15$ to describe the error distribution in a single simulation.

\begin{figure}
  \includegraphics[width=85mm]{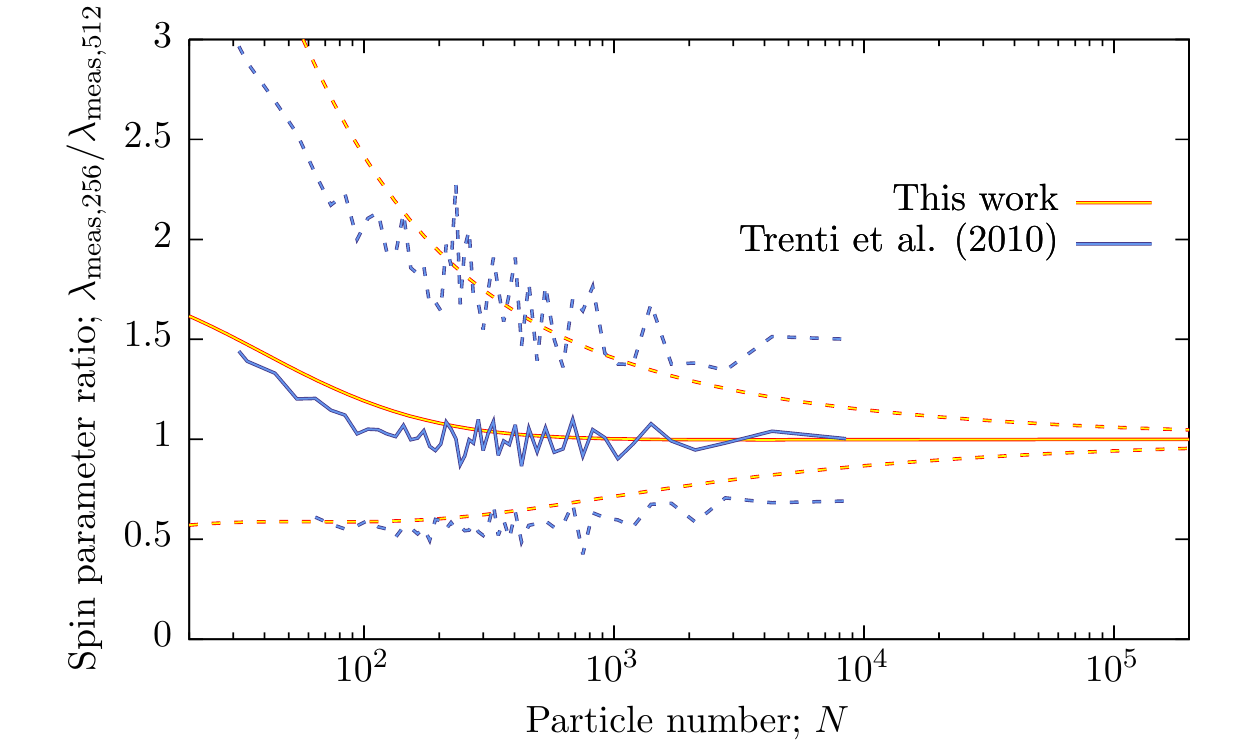}
  \caption{The median (solid lines) and symmetric $1\sigma$ intervals (dashed lines) of the distribution of the ratio of spin parameters in the medium box $256^3$ and $512^3$ particle simulations of \protect\cite{trenti_how_2010}. Blue lines show the results reported by \protect\cite[][their Fig.~9]{trenti_how_2010}, while yellow lines show the results of the model developed in this work when using the \protect\cite{trenti_how_2010} model for the error in halo mass as input.}
  \label{fig:spinErrorTrenti}
\end{figure}

\cite{trenti_how_2010} also measure errors in the spins of dark matter halos from their simulations. To assess if our model for errors in halo spin measurements is a viable description of the results of \cite{trenti_how_2010} we compute the mean and variance of the expected error distribution in their medium box $256^3$ and $512^3$ particle simulations using eqn.~(\ref{eq:errorTrenti}) as the error in halo particle number. We then compute the distribution of the ratio of spin parameters from the two distributions, modelling them as log-normal distributions. Finally, we plot the median and symmetric $1\sigma$ intervals of the resulting distribution to compare directly with the data of \cite{trenti_how_2010} as shown in Figure~\ref{fig:spinErrorTrenti}.

The result is a reasonable match to both the median and dispersion in spin parameter ratio, suggesting that our model is a good description of their results, particularly for the $N\ge300$ regime of interest here. At $N\approx 10^4$ our model seems to underpredict the error in spin measurement, although the \cite{trenti_how_2010} results are quite noisy in this regime. We note that a 10\% precision measurement of halo spin requires around $4\times 10^4$ particles in a halo.

We conclude that our model for the distribution of errors in spin parameters measured in N-body simulations, which was constrained to results from Poisson sampling of dark matter halos in the Millennium Simulation, also successfully describes the errors found in the study of \cite{trenti_how_2010} who considered how spin parameters change as the resolution of a simulation is increased.

\subsection{Modeling the spin distribution}

As we have shown, this simple model for the error distribution of halo spins measured in N-body simulations agrees quite well with the results of Monte Carlo experiments using cosmological N-body simulations, and reproduces the errors measured by \cite{trenti_how_2010}. Using this model we can construct a forward model for the spin distribution which will be measured in an N-body simulation, $p^\prime(\lambda^\prime)$, given some intrinsic (error-free) spin distribution, $p(\lambda)$. The parameters of the intrinsic spin distribution can then be constrained. Specifically,
\begin{equation}
 p^\prime(\lambda^\prime) = \frac{1}{N_{\rm h}} \int_{M_{\rm min}}^\infty {\rm d} M n(M) \int_0^\infty {\rm d}\lambda p(\lambda) g(\lambda^\prime|M,\lambda)
\end{equation}
where
\begin{equation}
N_{\rm h} =  \int_{M_{\rm min}}^\infty {\rm d} M n(M),
\end{equation}
$n(M)$ is the halo mass function, $M_{\rm min}$ is the minimum halo mass used in the N-body measurement of $p^\prime(\lambda)$, and $g(\lambda^\prime|M,\lambda)$ is the probability to measure a spin $\lambda^\prime$ in a halo of true mass $M$ and true spin $\lambda$. 

We choose to constrain the intrinsic spin distribution using the measurements of \Bett, specifically their ``TREEclean'' sample of halos. In constructing $g(\lambda^\prime|M,\lambda)$ for this purpose we boost the term corresponding to the halo energy in the spin-dependent error term to account for the fact that \Bett\ used at most 1,000 particles when estimating the potential energy of halos, although this has only a very small effect on our results.

For the intrinsic spin distribution we use the form proposed by \Bett:
\begin{equation}
 P(\log \lambda) = A \left( {\lambda \over \lambda_0} \right)^3 \exp \left[-\alpha \left( { \lambda \over \lambda_0 } \right)^{3/\alpha} \right],
 \label{eq:spinDistribution}
\end{equation}
where $\alpha$ and $\lambda_0$ are parameters to be determined\footnote{We also considered a log-normal distribution for the intrinsic spin, and assume this to be independent of halo mass. While such a distribution is better able to match the form of the N-body spin distribution after convolution with the effects of N-body particle noise than if such noise is ignored we find that it is still unable to provide an acceptable match to the N-body data.}.

To constrain these parameters we perform a differential evolution \MCMC\ simulation using the same approach as described in \cite{benson_mass_2016}. For the likelihood function we use
\mathchardef\mhyphen="2D
\begin{equation}
 \log \mathcal{L} = -\frac{1}{2}\sum_i \frac{\left(P_i^{\rm (model)}-P_i^{\rm (N\mhyphen body)}\right)^2}{V_i^{\rm (N\mhyphen body)}},
\end{equation}
where $P_i^{\rm (model)}$ is the model expectation for the distribution of measured spin parameter in the $i^{\rm th}$ bin of the \Bett\ distribution function, $P_i^{\rm (N\mhyphen body)}$ is the measured N-body distribution function in that bin from \Bett, and $V_i^{\rm (N\mhyphen body)}$ is the variance in that N-body measurement which we estimate assuming Poisson statistics and the total number (1,503,922) of halos in the TREEclean sample of \Bett. The two parameters $\alpha$ and $\lambda_0$ are assigned uniform priors on the ranges $(1.5,3.0)$ and $(0.025,0.055)$ respectively.

After the 48 \MCMC\ chains have converged we allow them to run for an additional 1121 steps. We find a correlation length of 8 steps in our chains. As such, the \MCMC\ simulation provides over 6,000 post-convergence independent samples from the posterior distribution of the parameters of the intrinsic spin distribution. 

\section{Results}\label{sec:results}

\begin{figure}
  \includegraphics[width=85mm]{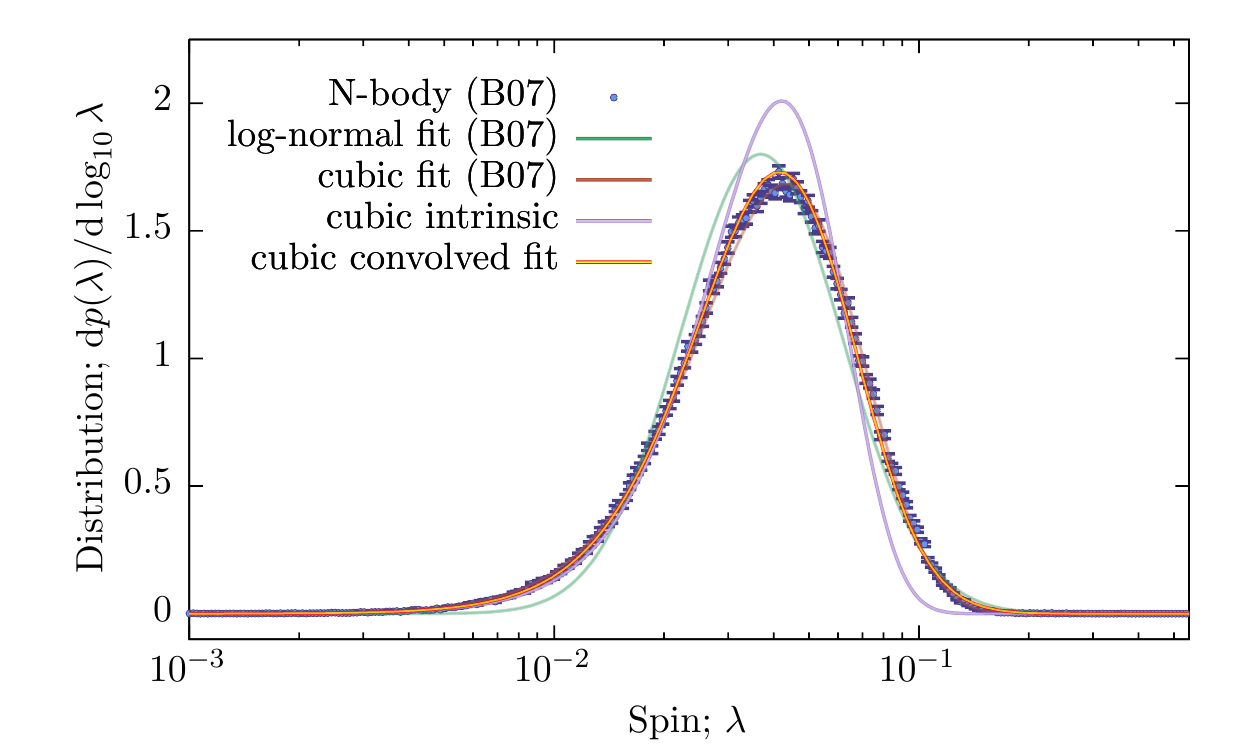}
  \caption{The distribution of spin parameters. Points indicate the distribution measured from the Millennium Simulation by \protect\Bett\ (we show only every fifth point from their histogram for clarity). The faint line indicates the best fit log-normal distribution found by \protect\Bett, while the red line indicates the best-fit ``cubic'' fit found by \protect\Bett. The yellow line indicates the best-fit to the N-body data found in this work, taking a log-normal distribution as the intrinsic, error-free model for the spin distribution. The purple line indicates that intrinsic, error-free distribution.}
  \label{fig:spinDistribution}
\end{figure}

Figure~\ref{fig:spinDistribution} shows the distribution of spin parameters reported by \Bett\ in their TREEclean sample (points with errorbars---we show only every fifth point for clarity), together with various analytic spin distribution functions. The green line is the best-fit log-normal distribution reported by \Bett\ and is a very poor match to the data (being unable to match the shape of the distribution at low spins), while the dark red line is the best fit distribution of the form given by equation~(\ref{eq:spinDistribution}) as reported by \Bett. The yellow line indicates the best fit distribution found in this work, and corresponds to the form given by equation~(\ref{eq:spinDistribution}) but convolved with the expected spin error distribution according to the model developed in \S\ref{sec:methods}. The purple line shows the corresponding intrinsic (i.e. unconvolved) spin distribution. It is apparent that the best fit distribution found in this work is a close match to the N-body measurements---in particular, we find a reduced $\chi^2=3.37$ for this model, which is somewhat worse than the value of $\chi^2=2.58$ reported by \Bett\ when fitting this same functional form without convolving with the particle noise error distribution. Both cases are formally poor fits, indicating that the functional form is not actually a good description of the N-body results.

\begin{figure}
 \newcommand{\triangledir}{.}
\renewcommand{\arraystretch}{0}
\setlength{\tabcolsep}{0pt}
\begin{tabular}{l@{}c@{}r@{}l@{}c@{}r@{}}
\multicolumn{3}{c}{\includegraphics[scale=1.0]{\triangledir/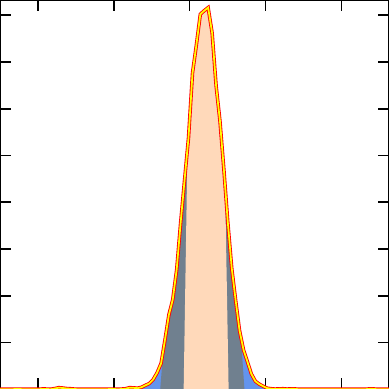}}&\multicolumn{3}{c}{\includegraphics[scale=1.0]{\triangledir/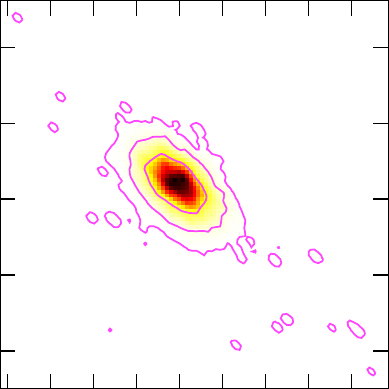}}\\
\multicolumn{1}{p{35.6923076923077pt}}{\raisebox{112pt-\widthof{\scriptsize x}-\widthof{\scriptsize $4.18 \times 10^{-2}$}}[0pt][0pt]{\rotatebox{90}{\scriptsize $4.18 \times 10^{-2}$}}}&\multicolumn{1}{p{35.6923076923077pt}}{\raisebox{112pt-\widthof{\scriptsize x}-\widthof{\scriptsize $\lambda_0$}}[0pt][0pt]{\rotatebox{90}{\scriptsize $\lambda_0$}}}&\multicolumn{1}{p{35.6923076923077pt}}{\raisebox{112pt-\widthof{\scriptsize x}-\widthof{\scriptsize $4.23 \times 10^{-2}$}}[0pt][0pt]{\rotatebox{90}{\scriptsize $4.23 \times 10^{-2}$}}}&\multicolumn{3}{c}{\includegraphics[scale=1.0]{\triangledir/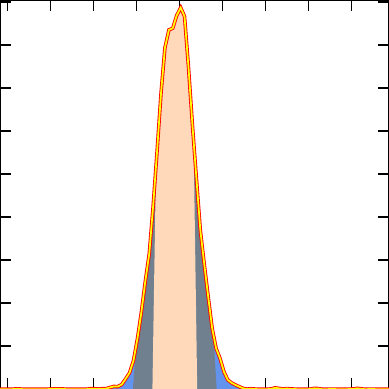}}\\
&&&\multicolumn{1}{p{35.6923076923077pt}}{\raisebox{-4pt-\widthof{\scriptsize $1.67$}}[-4pt][-4pt]{\rotatebox{90}{\scriptsize $1.67$}}}&\multicolumn{1}{p{35.6923076923077pt}}{\raisebox{-4pt-\widthof{\scriptsize $\alpha$}}[-4pt][-4pt]{\rotatebox{90}{\scriptsize $\alpha$}}}&\multicolumn{1}{p{35.6923076923077pt}}{\raisebox{-4pt-\widthof{\scriptsize $1.76$}}[-4pt][-4pt]{\rotatebox{90}{\scriptsize $1.76$}}}\\
\end{tabular}

 \vspace{0.9cm}
 \caption{Constraints on the parameters, $\alpha$ and $\lambda_0$ of the intrinsic spin distribution of the TREEclean sample of \protect\Bett. The off-diagonal panel shows the posterior distribution over both model parameters, while on-diagonal panels show the posterior distribution over individual model parameters. In the off-diagonal panel, colours show the probability density running from white (low probability density) to dark red (high probability density). Contours are drawn to enclose 99.7\%, 95.4\%, and 68.3\% of the posterior probability when ranked by probability density (i.e. the highest posterior density intervals). In on-diagonal panels the curve indicates the probability density. Shaded regions indicate the 68.3\%, 95.4\%, and 99.7\% highest posterior density intervals.}
 \label{fig:triangle}
\end{figure}

Figure~\ref{fig:triangle} shows the constraints obtained on the parameters of the spin distribution in equation~(\ref{eq:spinDistribution}). We find constraints of $\lambda_0$ $=(4.20190^{+0.00262}_{-0.00204}) \times 10^{-2}$ and $\alpha$ $=1.70918^{+0.00384}_{-0.00439}$ when marginalized over the other parameter. For comparison, \Bett\ found $\lambda_0 = 0.04326 \pm 0.000020$ and $\alpha = 2.509 \pm 0.0033$. Ignoring the effects of particle noise therefore results in $\lambda_0$ being overestimated by around 3\%. Given the huge statistical precision of the $1.5\times 10^6$ halo sample this corresponds to a bias in $\lambda_0$ of approximately $40\sigma$ when particle noise effects are ignored, while the shape parameter $a$ is biased by over $100\sigma$.

For our best fit parameters, the median measured spin is $0.0376$, while the median intrinsic spin is $0.03692$, again showing that ignoring particle noise effects results in typical spins being misestimated by around 2\% for this sample.

\section{Discussion}\label{sec:discussion}

N-body simulations are an invaluable tool for studying the non-linear regime of cosmological structure formation. As such simulations become ever larger in terms of particle number and volume, the statistical uncertainty (arising from finite numbers of halos for example) in measured properties is becoming very small. Consequently, it is now important to understand exactly what a measurement in an N-body simulation means, and how such should be interpreted. Any measurement will be affected by noise due to finite particle number. In some instances, such as halo spins discussed in this work, those effects can be significant in an absolute sense and can bias results.

We have shown how forward modeling of the effects of finite particle number allows constraints to be placed on the true, noise-free distribution of dark matter halo properties. Such an approach---providing the forward model is sufficiently accurate---will result in constraints which are independent of resolution, and which are unbiased. We have further demonstrated that such forward models can be validated through comparison with Monte Carlo experiments in which the particles of N-body halos are bootstrap resampled, and have shown that this bootstrapping approach itself gives results consistent with those obtained by running the same N-body simulation many times using different random sampling of the initial conditions (see Appendix~\ref{sec:nbodyStats}). While such a bootstrapping approach misses some sources of error (such as gravitational forces from subresolution structures; \citealt{trenti_how_2010}), our results show that this approach is nevertheless useful (and not prohibitively computationally expensive) in providing a general means for understanding particle noise errors in any measurement from N-body halos and for calibration of models of this noise.

In the specific case of halo spin measured from the TREEclean sample of \Bett\ we find that the parameter $\lambda_0$ is overestimated by 3\% if the effects of particle noise are ignored. Given the statistical precision to which this parameter can be constrained this corresponds to a bias of around $40\sigma$. It is worth noting that \Bett\ already limited their TREEclean samples to halos with $N>300$ particles in order to mitigate the effects of bias introduced by particle noise. We have shown that, even with this mitigation, a bias remains at a significant level. The effect is particularly important at very high spins. For $\lambda > 0.1$ the intrinsic spin distribution is very strongly suppressed relative to that measured directly from the simulation---almost all of these high-spin halos are in fact just the results of noise.

Halo spin is of particular importance for several astrophysical quantities, such as the sizes of galactic disks and for understanding intrinsic alignments of galaxies with large scale structure. In the case of galactic disk sizes, using the intrinsic spin distribution constrained in this work will result in galaxies being smaller in halos of given mass---although the effect is small (around 3\%) it is non-negligible compared to the accuracy of measured disk size distributions \citep[e.g.][]{shen_size_2003}. In the case of intrinsic alignments of galaxies with large scale structure, N-body simulations are often used to measure correlations \citep{kiessling_galaxy_2015}. While it is well-understood that a sufficient number of particles per halo is required to mitigate the effects of bias introduced by particle noise \citep{kiessling_galaxy_2015}, particle noise will also act to weaken any correlation between halo spin and large scale structure---this again could and should be forward modeled when estimating the effects of intrinsic alignments from N-body simulations.

Finally, the above discussion---specifically that particle noise-induced biases are around 3\%---applies to statistical properties of large ensembles of halos. For individual halos Figure~\ref{fig:spinErrorTrenti} clearly shows that errors on measurements of their spins can be much larger, around 100\% for $N=200$ halos and reaching 10\% only for halos containing $N=4\times 10^4$ particles. Many semi-analytic models of galaxy formation base their modeling of disk sizes on halo spin parameters. In many, these spin parameters are taken directly from N-body simulation measurements, including from halos resolved with only a small number of particles. For example, the semi-analytic model described by \citeauthor{croton_many_2006}~(\citeyear{croton_many_2006}; similar comments apply to \citealt{tecce_ram_2010,croton_semi-analytic_2016}), and many of its descendants, utilizes spins measured from the Millennium Simulation \citep{springel_simulations_2005} using halos containing as few as $N=20$ particles. These spins will be entirely dominated by particle noise and biased high at a very significant level. These errors and biases will  propagate, through modeling of disk sizes, into calculations of star formation rates, luminosities, and other properties of the model galaxies. We would recommend utilizing N-body-measured halo spins in semi-analytic models only for halos with $N\gtrsim 300$ particles, where the bias effect is at least sub-dominant (random errors will still be 
significant, but this is at least marginalized when comparing properties of populations of objects).

In summary, N-body simulations should be viewed as the results of a statistical experiment applied to a model of dark matter structure formation. When viewed in this way it is clear that determination of any quantity from such a simulation should be made through forward modeling of the effects of particle noise.

\section*{Acknowledgments}

We thank Peter Behroozi, Alina Kiessling, Yao-Yuan Mao, and Simon White for helpful conversations. The Millennium Simulation databases used in this paper and the web application providing online access to them were constructed as part of the activities of the German Astrophysical Virtual Observatory (GAVO). This research was supported in part by the National Science Foundation under Grant No. NSF PHY-1125915.

\bibliographystyle{mn2e}
\bibliography{spinDistributionAccented,spinDistributionExtra}

\begin{thebibliography}{31}
\expandafter\ifx\csname natexlab\endcsname\relax\def\natexlab#1{#1}\fi

\bibitem[{Bagla \& Padmanabhan(1997)}]{bagla_cosmological_1997}
Bagla J.~S., Padmanabhan T., 1997, Pramana, 49, 161

\bibitem[{Barnes \& Efstathiou(1987)}]{barnes_angular_1987}
Barnes J., Efstathiou G., 1987, ApJ, 319, 575

\bibitem[{Behroozi {et~al.}(2013)Behroozi, Wechsler, \&
  Wu}]{behroozi_rockstar_2013}
Behroozi P.~S., Wechsler R.~H., Wu H.-Y., 2013, ApJ, 762, 109

\bibitem[{Benson(2010)}]{benson_galaxy_2010}
Benson A.~J., 2010, Physics Reports, 495, 33

\bibitem[{Benson(2016)}]{benson_mass_2016}
---, 2016, ArXiv e-prints, 1610, arXiv:1610.01057

\bibitem[{Bett {et~al.}(2007)Bett, Eke, Frenk, Jenkins, Helly, \&
  Navarro}]{bett_spin_2007}
Bett P., Eke V., Frenk C.~S., Jenkins A., Helly J., Navarro J., 2007, MNRAS,
  376, 215

\bibitem[{Cole \& Lacey(1996)}]{cole_structure_1996}
Cole S., Lacey C., 1996, MNRAS, 281, 716

\bibitem[{Croton {et~al.}(2006)Croton, Springel, White, De~Lucia, Frenk, Gao,
  Jenkins, Kauffmann, {et~al.}}]{croton_many_2006}
Croton D.~J., Springel V., White S. D.~M., De~Lucia G., Frenk C.~S., Gao L.,
  Jenkins A., Kauffmann G., {et~al.}, 2006, MNRAS, 365, 11

\bibitem[{Croton {et~al.}(2016)Croton, Stevens, Tonini, Garel, Bernyk, Bibiano,
  Hodkinson, Mutch, {et~al.}}]{croton_semi-analytic_2016}
Croton D.~J., Stevens A. R.~H., Tonini C., Garel T., Bernyk M., Bibiano A.,
  Hodkinson L., Mutch S.~J., {et~al.}, 2016, ApJS, 222, 22

\bibitem[{Davis {et~al.}(1985)Davis, Efstathiou, Frenk, \&
  White}]{davis_evolution_1985}
Davis M., Efstathiou G., Frenk C.~S., White S. D.~M., 1985, ApJ, 292, 371

\bibitem[{Doroshkevich(1970)}]{doroshkevich_spatial_1970}
Doroshkevich A.~G., 1970, Astrophysics, 6, 320

\bibitem[{Fall \& Efstathiou(1980)}]{fall_formation_1980}
Fall S.~M., Efstathiou G., 1980, MNRAS, 193, 189

\bibitem[{Gottl\"ober \& Yepes(2007)}]{gottlober_shape_2007}
Gottl\"ober S., Yepes G., 2007, ApJ, 664, 117

\bibitem[{Hartigan \& Hartigan(1985)}]{hartigan_dip_1985}
Hartigan J.~A., Hartigan P.~M., 1985, Ann. Statist., 13, 70

\bibitem[{Hoyle(1949)}]{hoyle_origin_1949}
Hoyle F., 1949, in Problems in {Cosmical} {Aerodynamics}, Proceedings of the
  {Symposium} on the {Motion} of {Gaseous} {Masses} of {Cosmical} {Dimensions},
  Central Air Documents Officem Ohio

\bibitem[{Kiessling {et~al.}(2015)Kiessling, Cacciato, Joachimi, Kirk,
  Kitching, Leonard, Mandelbaum, Sch\"afer, {et~al.}}]{kiessling_galaxy_2015}
Kiessling A., Cacciato M., Joachimi B., Kirk D., Kitching T.~D., Leonard A.,
  Mandelbaum R., Sch\"afer B.~M., {et~al.}, 2015, Space Science Reviews, 193,
  67

\bibitem[{Lee {et~al.}(2016)Lee, Primack, Behroozi, Rodr\'iguez-Puebla,
  Hellinger, \& Dekel}]{lee_properties_2016}
Lee C.~T., Primack J.~R., Behroozi P., Rodr\'iguez-Puebla A., Hellinger D.,
  Dekel A., 2016, MNRAS

\bibitem[{Macci\`o {et~al.}(2007)Macci\`o, Dutton, van~den Bosch, Moore,
  Potter, \& Stadel}]{maccio_concentration_2007}
Macci\`o A.~V., Dutton A.~A., van~den Bosch F.~C., Moore B., Potter D., Stadel
  J., 2007, MNRAS, 378, 55

\bibitem[{Mo {et~al.}(1998)Mo, Mao, \& White}]{mo_formation_1998}
Mo H.~J., Mao S., White S. D.~M., 1998, MNRAS, 295, 319

\bibitem[{Peebles(1969)}]{peebles_origin_1969}
Peebles P. J.~E., 1969, ApJ, 155, 393

\bibitem[{Porciani {et~al.}(2002)Porciani, Dekel, \&
  Hoffman}]{porciani_testing_2002}
Porciani C., Dekel A., Hoffman Y., 2002, MNRAS, 332, 325

\bibitem[{Poveda-Ruiz {et~al.}(2016)Poveda-Ruiz, Forero-Romero, \& Mu\~noz
  Cuartas}]{poveda-ruiz_quantifying_2016}
Poveda-Ruiz C.~N., Forero-Romero J.~E., Mu\~noz Cuartas J.~C., 2016, ApJ, 832,
  169

\bibitem[{Rodr\'iguez-Puebla {et~al.}(2016)Rodr\'iguez-Puebla, Behroozi,
  Primack, Klypin, Lee, \& Hellinger}]{rodriguez-puebla_halo_2016}
Rodr\'iguez-Puebla A., Behroozi P., Primack J., Klypin A., Lee C., Hellinger
  D., 2016, MNRAS, 462, 893

\bibitem[{Shen {et~al.}(2003)Shen, Mo, White, Blanton, Kauffmann, Voges,
  Brinkmann, \& Csabai}]{shen_size_2003}
Shen S., Mo H.~J., White S. D.~M., Blanton M.~R., Kauffmann G., Voges W.,
  Brinkmann J., Csabai I., 2003, MNRAS, 343, 978

\bibitem[{Springel(2005)}]{springel_cosmological_2005}
Springel V., 2005, MNRAS, 364, 1105

\bibitem[{Springel {et~al.}(2005)Springel, White, Jenkins, Frenk, Yoshida, Gao,
  Navarro, Thacker, {et~al.}}]{springel_simulations_2005}
Springel V., White S. D.~M., Jenkins A., Frenk C.~S., Yoshida N., Gao L.,
  Navarro J., Thacker R., {et~al.}, 2005, Nature, 435, 629

\bibitem[{Tecce {et~al.}(2010)Tecce, Cora, Tissera, Abadi, \&
  Lagos}]{tecce_ram_2010}
Tecce T.~E., Cora S.~A., Tissera P.~B., Abadi M.~G., Lagos C. D.~P., 2010,
  MNRAS, 408, 2008

\bibitem[{Trenti {et~al.}(2010)Trenti, Smith, Hallman, Skillman, \&
  Shull}]{trenti_how_2010}
Trenti M., Smith B.~D., Hallman E.~J., Skillman S.~W., Shull J.~M., 2010, ApJ,
  711, 1198

\bibitem[{White(1984)}]{white_angular_1984}
White S. D.~M., 1984, ApJ, 286, 38

\bibitem[{Zhang {et~al.}(2009)Zhang, Yang, Faltenbacher, Springel, Lin, \&
  Wang}]{zhang_spin_2009}
Zhang Y., Yang X., Faltenbacher A., Springel V., Lin W., Wang H., 2009, ApJ,
  706, 747

\bibitem[{Zjupa \& Springel(2017)}]{zjupa_angular_2017}
Zjupa J., Springel V., 2017, MNRAS, 466, 1625

\end{thebibliography}

\appendix

\section{Validating Bootstrapping of N-body Halo Particles to Estimate Errors}\label{sec:nbodyStats}

In this work \citep[also in][]{poveda-ruiz_quantifying_2016,benson_mass_2016} we use a bootstrap resampling procedure to estimate the errors on quantities measured from N-body simulations. Specifically, given an N-body representation of a dark matter halo we bootstrap resample the particles (with replacement) a large number of times, each time measuring the quantity of interest, and use the ensemble of measurements to estimate the variance in the measured property. Before describing how we validate this assumption we reiterate that this approach does not capture the full error budget of the N-body simulation, as it neglects the effects of missing subresolution structure \citep{trenti_how_2010}. Nevertheless, it provides a valuable experimental technique to assess the effects of errors, and to provide measurements to which models of those errors can be calibrated and validated.

This bootstrapping procedure is an attempt to understand how the finite number of random samples (i.e. particles) from the dark matter distribution function affect our ability to measure the true value of some quantity. The correct way to assess this particle-induced variance would be to repeat the N-body simulation a large number of times, each with the same initial density field, but with that density field sampled by a different randomly chosen ensemble of particles each time. This approach avoids possible failings of the bootstrapping approach (such as the fact that the particles may not represent a Poisson process sampling of the dark matter distribution function due to the way initial conditions are constructed, and because the particles are also the sources of gravitational potential).

In this appendix we therefore carry out precisely this experiment, and then compare the results to models of the error distribution which assume Poisson sampling and which were previously calibrated to bootstrap estimates of the variance in halo mass \citep{benson_mass_2016} in order to validate the bootstrapping approach.

To generate initial conditions for our cosmological N-body simulation we construct 1024 realizations of glass distributions in a 50~Mpc$/h$ box consisting of $64^3$ particles using Gadget2 \citep{springel_cosmological_2005}. The glass is created by running for 256 steps, by which time the rms particle movement is no longer decreasing significantly. We check that the variance in the glass is sub-Poissonian---for example, in cubic cells of volume 1~Mpc$^3$ the mean number of particles is $2.097$. For a Poisson distribution of particles the variance in the number of particles in such cells is also $2.097$, but for our glass we measure a variance of $0.7708$.

For each glass, we generate initial conditions at $z=63$ using N-GenIC \citep{springel_simulations_2005} for a $(\Omega_{\rm M},\Omega_\Lambda,\Omega_{\rm b},H_0,\sigma_8)=(0.3071,0.6929,0.04820,67.7~\hbox{km/s/Mpc},0.8228)$ cosmology using the same random seed such that the amplitudes and phases of modes in the density field are identical in each case. Importantly though, since we use an independently constructed glass distribution of particles in each case, the sampling of modes by particles is different in each realization.

Each realization is then evolved to $z=0$ using Gadget2 with a softening length of 7.8125~kpc ($1/50$ of the mean interparticle separation), after which we use the Rockstar halo finder \citep{behroozi_rockstar_2013} with a friends-of-friends linking length of $b=0.28$ and a minimum group size of $N=20$ to extract a halo catalog. We then matched halos\footnote{We did not consider subhalos.} between pairs of realizations by selecting those whose centres are offset by the smallest fraction of the minimum of the two virial radii, excluding cases where no match is found within the smaller of the two virial radii. We begin from the most massive halos and work down to lower masses. For halos where we are able to match across 256 or more realizations (such that we have sufficient halos from which to compute a variance) we then compute the mean and variance in the recovered halo mass. As such, this variance correctly incorporates the effects of initial particle distribution, the effects of the particles being carriers of the density field as well as tracers of it, and any possible vagaries of the halo finder.

\begin{figure*}
 \begin{tabular}{cc}
 \vspace{-5mm}\ifthenelse{\boolean{lores}}{\includegraphics[width=80mm]{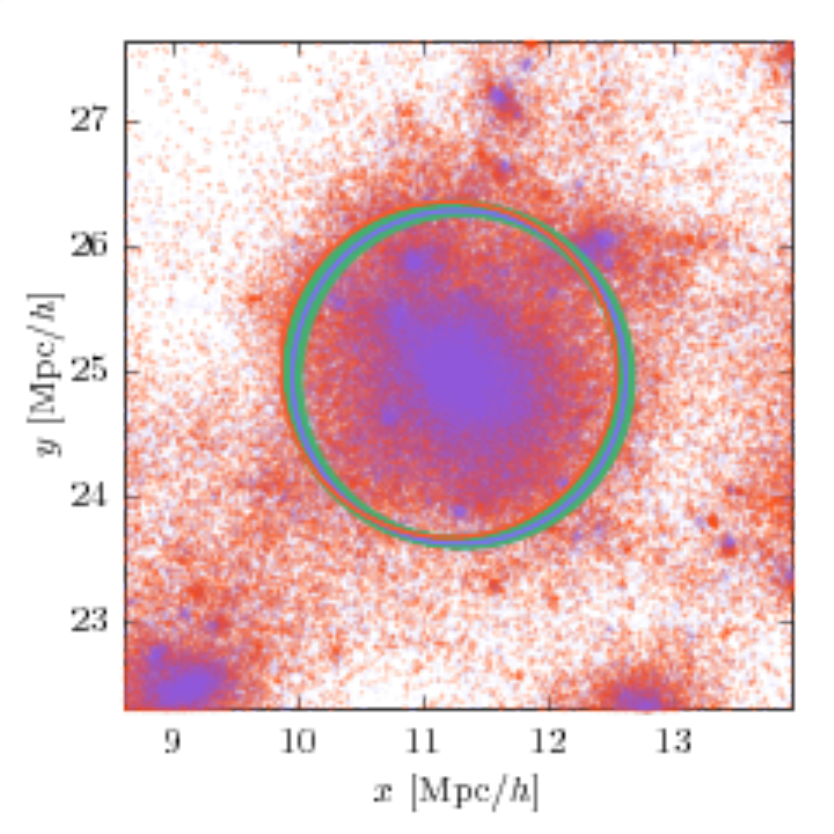}}{\includegraphics[width=80mm]{singleHaloRealizationsHighMass.pdf}} & 
 \includegraphics[width=80mm]{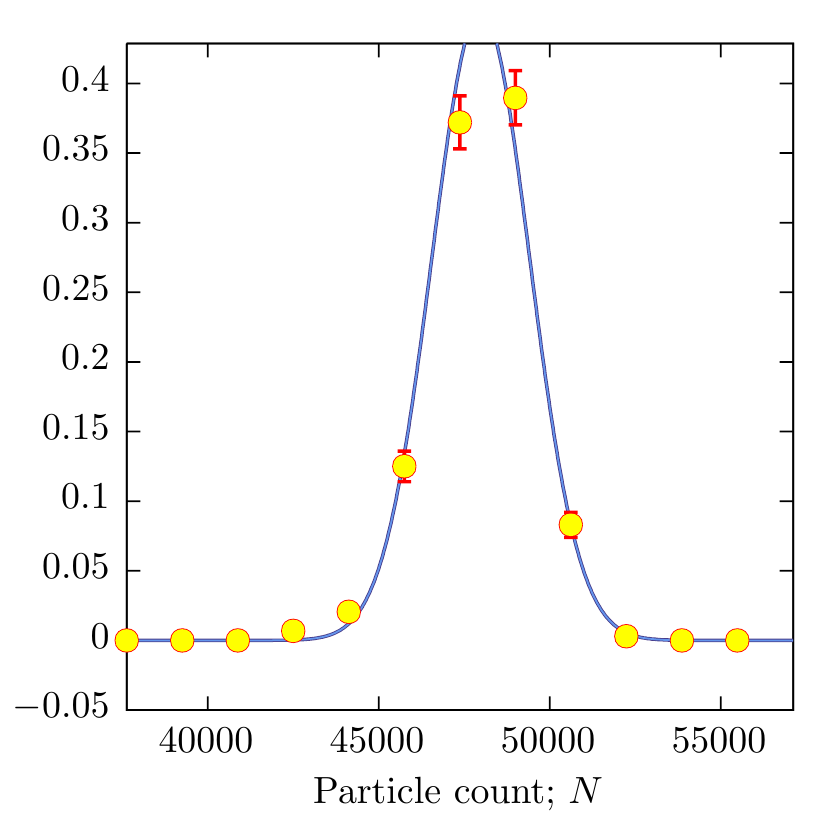} \\
 \vspace{-5mm}\includegraphics[width=80mm]{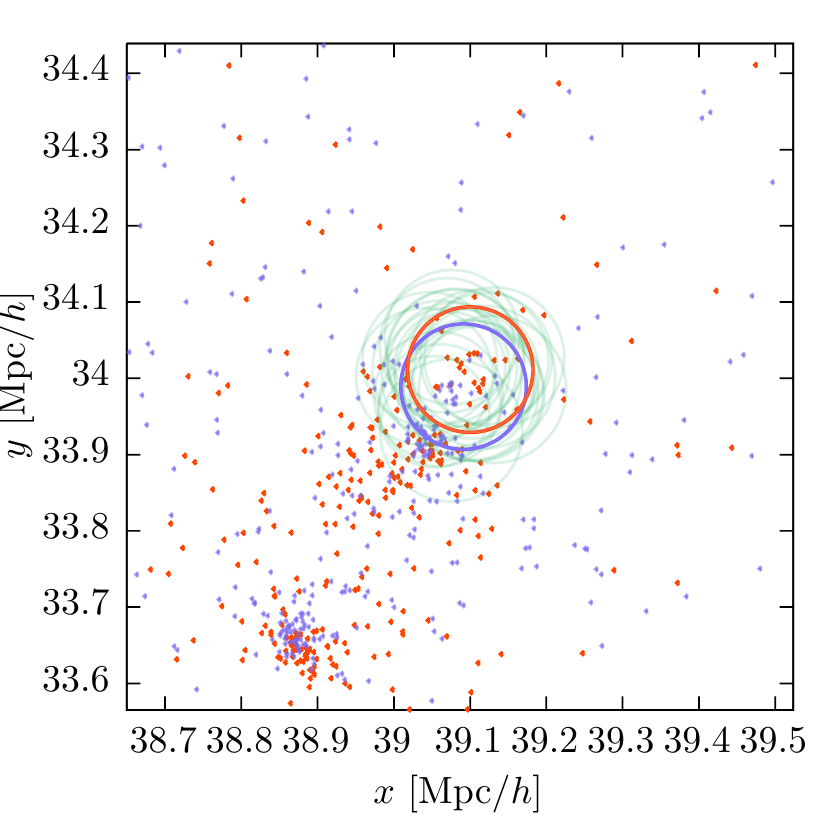} & 
 \includegraphics[width=80mm]{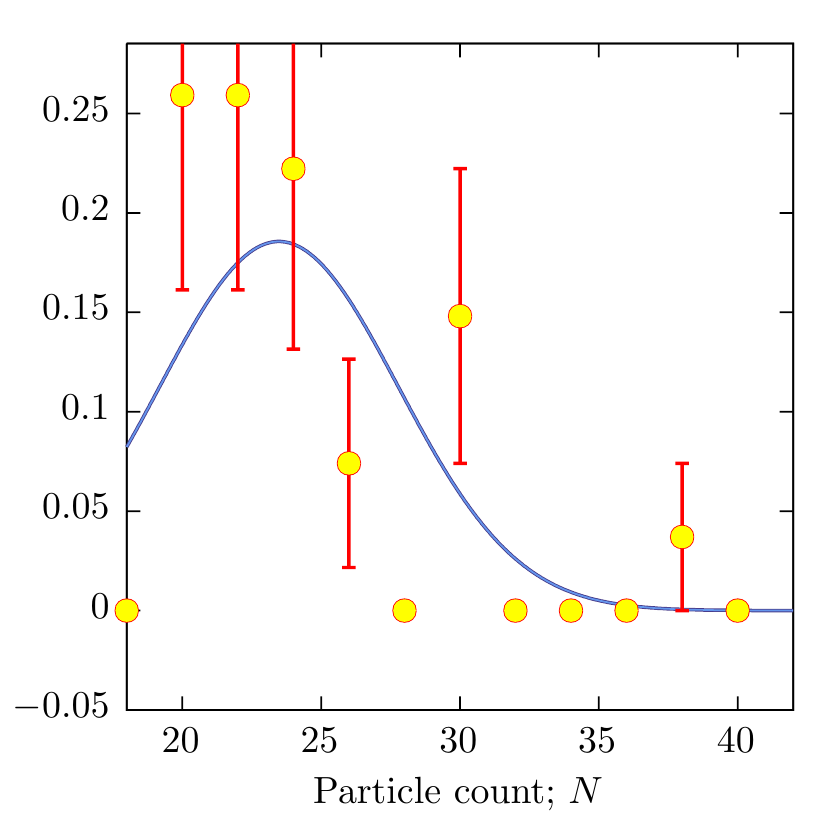} \\
 \vspace{-5mm}\ifthenelse{\boolean{lores}}{\includegraphics[width=80mm]{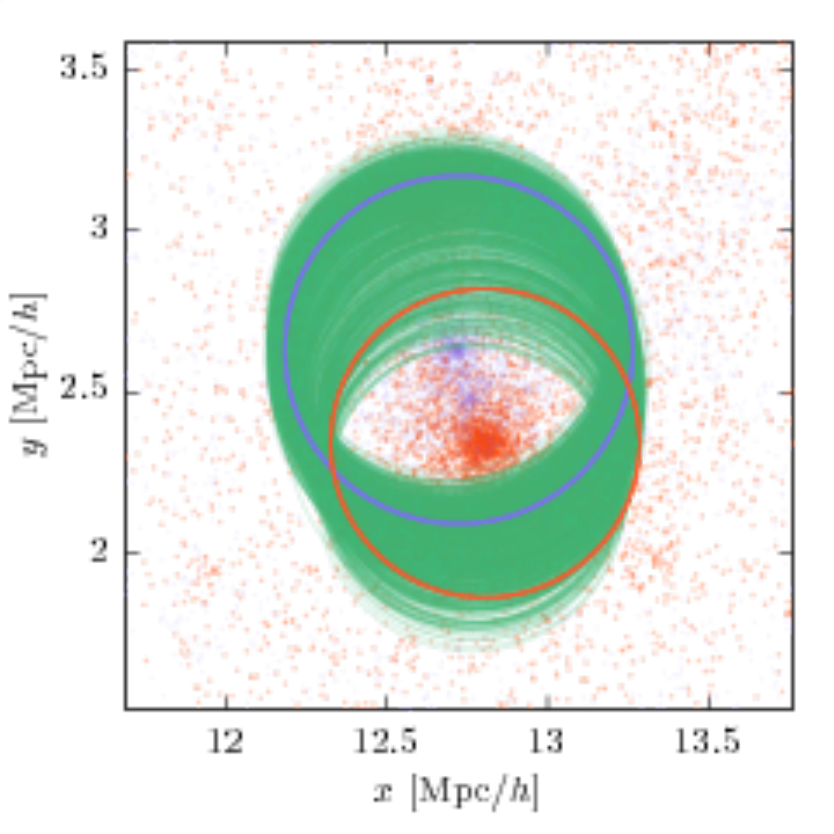}}{\includegraphics[width=80mm]{singleHaloRealizationsOutlier.pdf}} & 
 \includegraphics[width=80mm]{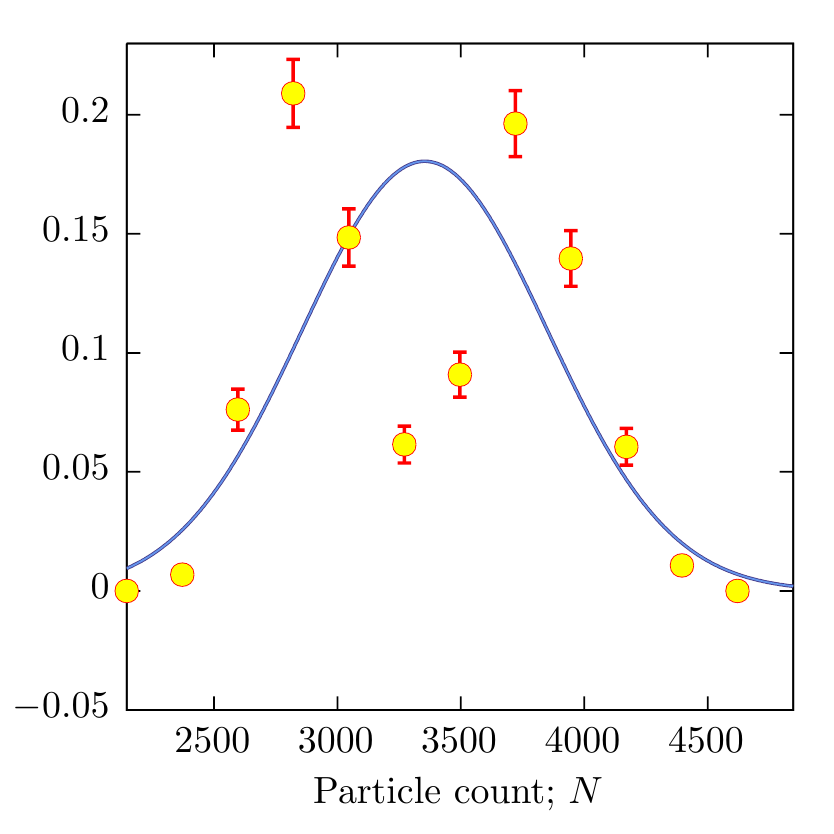} \\
 \end{tabular}
 \caption{Realizations of single halos. Circles indicate the centre and virial radius of each realization of a single halo in our simulations. Blue/red circles highlight the realizations in which the halo centre is closest to/farthest from the mean position of the halo over all realizations. Blue/red points indicate particles from those same two realizations. The top row indicates one of the highest mass halos, the middle row a very low mass halo, and the bottom row shows a halo flagged as an outlier as described in the text.}
 \label{fig:singleHaloRealizations}
\end{figure*}

Figure~\ref{fig:singleHaloRealizations} shows realizations of individual halos (left column) along with histograms of the number of particles in each FoF halo (right column). Green circles indicate the centres and virial radii of the halo as determined from a subsample of all N-body realizations. The blue circle indicates the realization of the halo whose center lies closest to the mean centre over all realizations, while the red circle indicates the realization of the halo which lies farthest from that mean centre. The red and blue points show the particle distribution corresponding to these two extreme realizations. Note that the blue points are substantially sub-sampled to avoid completely washing out the red points. The top row indicates one of the highest mass halos, the middle row a very low mass halo, and the bottom row shows a halo whose mass histogram is judged to have significant non-unimodality via the statistic of \cite{hartigan_dip_1985} (using $p<0.05$ to indicate the presence of multiple modes). In the right hand column the blue curve shows the best fitting normal distribution for comparison.

For the high-mass halo ($\langle N \rangle=47,986$, top row), we find very regular behaviour. There is some shifting in the position of the centre and radius between realizations. The blue realization (closest to the mean halo centre) lies well within the distribution of green circles, while the extreme case (red circle) is not too far offset (a small fraction of the virial radius). The histogram of particle count in the halo is consistent with a normal distribution.

In the low-mass halo ($\langle N \rangle=23.5$, middle row) the results are, not surprisingly, much noisier. In any given realization, the halo centre can be offset by a substantial fraction of the virial radius from the mean centre position. The histogram of halo mass is also noisy, but consistent with being drawn from a normal distribution (once we account for the fact that bins with $N\le 20$ are empty due to the limit imposed by the group finder).

In the outlier case ($\langle N \rangle =3,354$, bottom row) it is very clear from the mass histogram that we have a bimodal distribution of particle count, and it is entirely inconsistent with being drawn from a normal distribution. The masses of the two modes differ by around 30\%. Considering the particle distribution there is a very clear difference between the blue (closest to average) and red (farthest from average) cases, with the centres of the halos being distinctly offset from each other. This significant offset, in a halo which is well-resolved, arises entirely from the sampling of the initial density field of the simulation by different sets of randomly selected particles. In the case of this outlier (and many of the others which we examined) it is clear that there are two halos present which are in the process of merging. In some realizations the halos have approached sufficiently closely that they are linked together by the friends-of-friends algorithm into a single halo, while in other realizations the halos are still sufficiently far apart that they are identified as distinct halos.

\begin{figure*}
 \begin{tabular}{cc}
  \includegraphics[width=85mm]{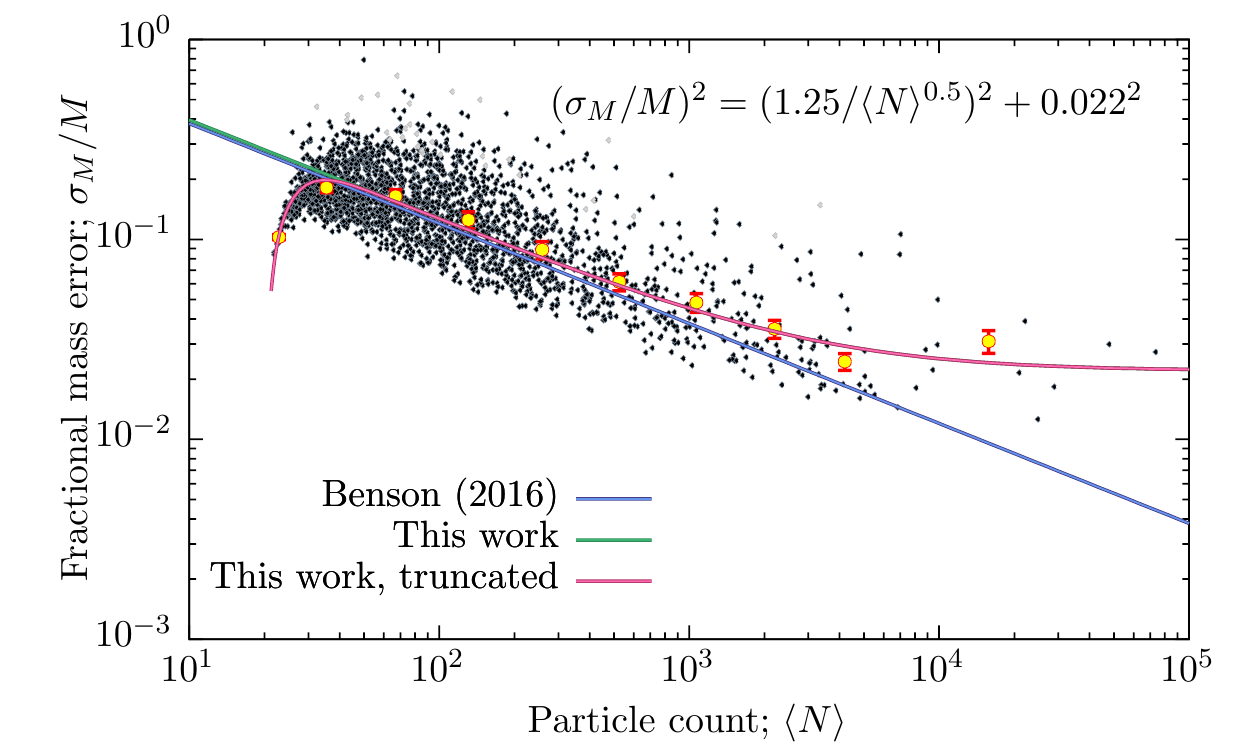} &
  \includegraphics[width=85mm]{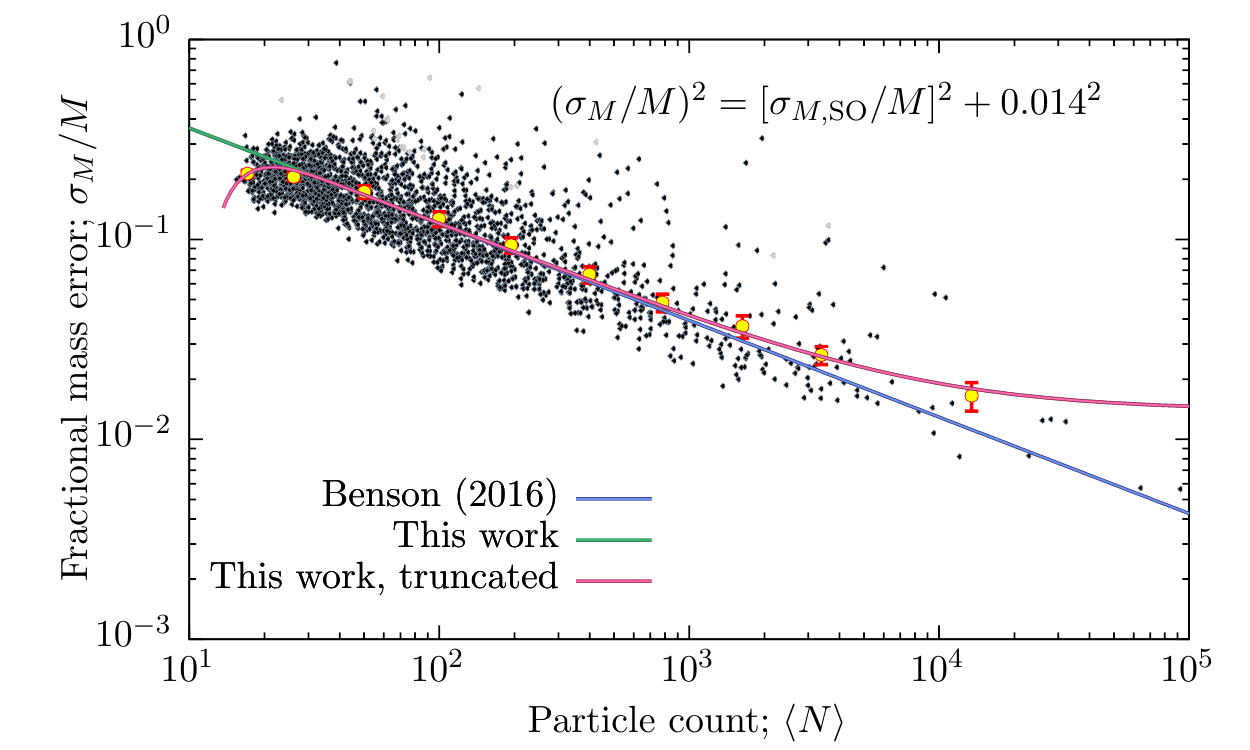}
 \end{tabular}
 \caption{Black points show fractional root-variances in halo masses as a function of particle number determined from 1024 cosmological N-body simulations each sharing the same initial density field, but sampled with independent, random sets of particles. The left-hand panel shows results for a friends-of-friends group finder, while the right-hand panel shows results for spherical overdensity group finder. Grey points indicate high-variance halos which are removed from the sample due to their mass histograms showing significant non-unimodality. Yellow circles indicate the mean fractional error, computed in bins at least a factor of 2 in width, or wide enough to contain 20 halos (whichever is larger). The blue line in each panel indicates the error model from \protect\cite{benson_mass_2016}, while the green line indicates the model fit to the yellow points (the form of which is reported in each panel). The purple line shows the same model as the green line but accounts for the biased introduced by the 20 particle minimum required by the group finder to identify halos.}
 \label{fig:nbodyErrors}
\end{figure*}

Figure~\ref{fig:nbodyErrors} shows the resulting root-variances in halo mass as a function of particle number for both friends-of-friends (FoF; linking length of $b=0.28$) and spherical overdensity (SO; density threshold of $\Delta = 200 \bar{\rho}$ where $\bar{\rho}$ is the mean density of the simulation) estimates of the halo mass. In each panel we plot the measured fractional mass error for each halo as a function of the mean number of particles in the halo as black points. Grey points indicate high-variance halos (those with root-variances exceeding twice the model expectation) which are removed from the sample due to their mass histograms showing significant non-unimodality (as judged by the statistic of \cite{hartigan_dip_1985} with $p<0.05$). Yellow circles indicate the mean fractional error, computed in bins at least a factor of 2 in width, or wide enough to contain 20 halos (whichever is larger). The blue line in each panel indicates the error model from \cite{benson_mass_2016}, while the green line indicates the model fit to the yellow points (the form of which is reported in each panel). Rockstar was configured to keep only halos containing 20 or more particles. As such, at low particle count our measurements become biased (as the distribution of halo particle counts is truncated below 20). To model this effect we assume that the particle count at each mass is a random variable drawn from a normal distribution with mean equal to the mean particle count, and variance taken from our error model. We then compute the mean and variance of this distribution when truncated below 20. The results are shown by the purple line. In the case of the FoF halo finder this approach accurately describes the behaviour at low masses. As such, we fit the parameters of our error model to the yellow points with this truncation modification in place. For the SO halo finder the truncation is more complicated, as the reported SO mass can lie below 20 particles providing the FoF mass (from which the original selection was made) is 20 or greater. The truncation is therefore not as extreme. We therefore adopt the same model but allow the cut-off point to be a free parameter, finding that a truncation below 12 particles describes the result well. 

Additionally, in both FoF and SO cases we see clear evidence that the error relation changes slope at high particle number. This is due to the fact that, in halos containing greater than $10^4$ particles, the {\sc Rockstar} halo finder uses a random subsample of $10^4$ particles when performing the friends-of-friends step to speed up computation (Behroozi, private communication). The effect of this optimization can be modeled by assuming an additional, constant error term (added in quadrature to the mass-dependent term). As a result, halo masses are never measured in N-body simulations to better than 1--2\%, even when very well resolved. This optimization can be switched off in {\sc Rockstar}, but we choose here to run {\sc Rockstar} in its default configuration---the presence of a constant fractional error at high particle number is easily modelled. Since the Millennium Simulation halo catalogs were constructed using a different halo finder, which does not have this same optimization, we do not include the constant error terms when modelling results from the Millennium Simulation.

We experimented with allowing the exponent of the $\langle N\rangle$ term in the FoF model to vary, but found it did not improve the fit sufficiently compared to a fixed exponent of $0.5$ (which is at least heuristically motivated by Poisson noise) to justify allowing this term to vary.

These results are largely consistent with those found in \cite{benson_mass_2016}, which assume that particles in N-body halos represent a Poisson process sampling of the dark matter distribution and which are consistent with estimates based on bootstrapping, justifying the bootstrapping approach taken there, with only a small recalibration of the amplitude of the FoF error model, and the inclusion of a constant error term for high mass halos\footnote{Neither of these would have significant effect on the results of \protect\cite{benson_mass_2016}, which used SO halos (and so is unaffected by the recalibration of the FoF model), and for which the effects of errors was dominated by low mass ($\sim300$ particle) halos for which the SO model without a constant term works very well.}. It is particularly interesting to note that the simple, parameter free model for errors in SO halos derived by \cite{benson_mass_2016} accurately describes the measured errors from 30 to approximately $10^4$ particles, beyond which the constant error term dominates.

\end{document}